\documentclass[reprint,
superscriptaddress,
 amsmath,amssymb,
 aps,
prl,
]{revtex4-2}

\usepackage{graphicx}% Include figure files
\usepackage{subfigure}
\usepackage{dcolumn}% Align table columns on decimal point
\usepackage{bm}% bold math
\usepackage{amsmath}
\usepackage{amsthm}
\usepackage{xcolor}
\usepackage{ulem}
\usepackage{pifont}
\usepackage{textcomp}
\usepackage[colorlinks, linkcolor = blue, anchorcolor = blue, citecolor = blue]{hyperref}
\usepackage{amsfonts}
\usepackage{dsfont}

\usepackage{graphicx,accents}

\makeatother
\newtheorem{thm}{Theorem}

\newcommand{\N}{\mathcal{N}}
\newcommand{\M}{\mathcal{M}}
\newcommand{\B}{\mathcal{B}}
\newcommand{\calO}{\mathcal{O}_{\vec{\mu}}}
\newcommand{\T}{\mathcal{T}}
\newcommand{\tr}{\text{Tr}}
\newcommand{\Q}{Q^{(1)}}

\begin{document}

\title{Super-additivity of quantum capacity in simple channels  }

\date{\today}% It is always \today, today,
%  but any date may be explicitly specified

%%%%% author %%%%%%%%%%%%%%%%%%%%%%%%%%%%%%%%%%%%

\author{Zhen Wu}
\affiliation{School of Mathematical Sciences, MOE-LSC, Shanghai Jiao Tong University, Shanghai, 200240, China}
\author{Qi Zhao}
\affiliation{QICI Quantum Information and Computation Initiative, Department of Computer Science, School of Computing and Data Science, The University of Hong Kong, Hong Kong}
\email{zhaoqi@cs.hku.hk}
\author{Zhihao Ma}
\affiliation{School of Mathematical Sciences, MOE-LSC, Shanghai Jiao Tong University, Shanghai, 200240, China}
\affiliation{Shanghai Seres Information Technology Co., Ltd, Shanghai 200040, China}
\affiliation{Shenzhen Institute for Quantum Science and Engineering, Southern University of Science and Technology, Shenzhen 518055, China}
\email{mazhihao@sjtu.edu.cn}
\begin{abstract}
    The super-additivity of quantum channel capacity is an important feature of quantum information theory different from classical theory, which has been attracting attention. Recently a special channel called ``platypus channel'' exhibits super-additive quantum capacity when combined with qudit erasure channels. Here we consider the ``generalized platypus channel'', prove that it has computable channel capacities, such as both private and classical capacity equal to $1$, and in particular, the generalized platypus channel still displays the super-additivity of quantum capacity when combined with qudit erasure channels and multilevel amplitude damping channels respectively.
\end{abstract}

\maketitle

\section{Introduction}

Shannon's information theory~\cite{Shannon} establishes the mathematical foundation for transmitting classical information through classical channels, and defines a key quantity: \emph{channel capacity} $C_\text{Shan}$, which is the maximum rate of classical information that can be reliably transmitted using a given classical channel. He proved the capacity of a given channel is the maximum mutual information between the input and output system of the channel taking over all possible input probability distribution. Shannon channel capacity, which is at the heart of the development of communication systems, has an important property: \emph{additivity}, which implies that the use of two classical channels in parallel has the same capacity as using them independently, that is $C_{\text{Shan}}(\N_1\times \N_2) = C_{\text{Shan}}(\N_1) + C_{\text{Shan}}(\N_2)$ for any two classical channels $\N_1$ and $\N_2$. However, the physical world is governed by quantum theory, prompting the development of quantum information theory~\cite{Wilde,WatrousTQI} as a more generic framework from its classical counterpart, e.g., data can be encoded into the microscopic particles of a quantum system as quantum information and transmitted by quantum channels.

Nevertheless, since quantum mechanics allows the existence of spatial correlations, such as entanglement, quantum information theory demonstrates qualitative distinctions from classical theory, which are prominently manifested in the definition of quantum channel capacities for a generic quantum channel $\mathcal{B}$. For instance, when transmitting quantum information, the maximum reliable transmission rate that can be achieved using a quantum channel $\mathcal{B}$ in a single pass is defined as its coherent information $\Q(\B)$. Whereas, if there are $n$ quantum channels $\{\B_i\}_{i=1}^n$ combined in parallel as a new channel $\N := \B_1\otimes \cdots \otimes \B_n$ to transmit quantum information, the maximum error-free rate that can be achieved by $\N$ is likely to exceed that of using these channels independently, i.e. $\Q(\N) > \sum_{i=1}^n \Q(\B_i)$ which is called \emph{super-additivity}. Therefore, the definition of quantum capacity $Q$ of a quantum channel $\B$, which represents the maximum rate of error-free quantum information transmission without assistance, needs to take into account super-additivity and is given by the following regularized expression~\cite{LSD1,LSD2,LSD3}
\begin{equation}\label{RegularizedExpresseion}
\begin{aligned}
    Q(\B) = \lim_{n\rightarrow \infty} \frac{Q^{(1)}(\B^{\otimes n})}{n}\,. \\
\end{aligned}
\end{equation}
This regularization becomes necessary, since for every integer $n_0 \in \mathbb{N}$, there exists a quantum channel $\N_{n_0}$ so that $\Q(\N_{n_0}^{\otimes m}) = 0$ for all $m \leq n_0$ yet $Q(\N_{n_0})>0$~\cite{UnboundedQ,UnboundP}. 
%\qi{from this sentence, I can not understand why it is different with the classical one, in classical they don't need to define asymptotic one?   what is $n_0$?}
Moreover, quantum information theory meticulously defines different quantum channel capacities according to specific communication tasks, for example, in addition to quantum capacity $Q$ mentioned above, there are the classical capacity $C$~\cite{classcapaHolevo,HSw1,HSw2} which quantifies the highest rate of a quantum channel transmitting classical information with error vanishing, and the private capacity $P$~\cite{Privacy,LSD3} of a quantum channel is about its capability for quantum cryptography~\cite{BB84}. Both the private and classical capacity require a similar regularized expression as Eq.~\eqref{RegularizedExpresseion}, which makes it very hard to evaluate the capacities of a generic quantum channel. In fact, even computing $\Q$ for a general quantum channel also requires non-convex optimization over infinite states, which is also not feasible.

The super-additivity can be divided into two categories based on the violation of weak and strong additivity. Let's take coherent information as an example. Formally, a quantum channel is said to have \emph{weakly additive} coherent information if $\Q(\N^{\otimes n}) = n\Q(\N)$ for any $n\in \mathbb{N}$, which leads a convenient formula for its quantum capacity: $Q(\N) = \Q(\N)$. Extending to multi-channel scenarios, the inequality of coherent information
\begin{equation}\label{SuperAdd}
    \Q(\B_1\otimes\B_2) \geq \Q(B_1) + \Q(\B_2)
\end{equation}
generally holds. Fixed the quantum channel $\mathcal{B}_1$, once the equality is saturated for all quantum channels $\mathcal{B}_2$, we call $\B_1$ has \emph{strongly} additive coherent information $\Q$ and hence the quantum capacity of $\B_1$ is also strongly additive: $Q(\mathcal{B}_1\otimes\mathcal{B}_2) = Q(\mathcal{B}_1) + Q(\mathcal{B}_2)$. Such inequality can be strict, in this situation, we call the coherent information $\Q$ of $\B_1$ and $\B_2$ are \emph{super-additive}, which implies that using $\B_1$ and $\B_2$ in parallel can transmit more information than use them separately. Notably, whereas additivity holds universally in classical theory, quantum systems demonstrate striking violations of both weak and strong additivity, such as, violating weak additivity~\cite{SuperAddIcPauli,SuperAddIcPauli2,DephrasureChannel,ErrorThresholdsPauli,IEEEDeep,geneticalgorithms} and strong additivity~\cite{IEEEPos,Filippov_2021,Singular} of coherent information $\Q$; and violation weak additivity of private information $P^{(1)}$ introduced below~\cite{StructuredCodes,SuperAddIp1,UnboundP}; violating the strong additivity of Holevo capacity $C^{(1)}$~\cite{Hastings2009}, and even quantum capacity~\cite{superactivation,Superact2} as well as private capacity~\cite{SuperAddP2,SuperAddP} while the super-additivity of classical capacity $C$ is still open. In contrast, few types of channels are known to have additive channel capacity, for instance, the quantum and private capacity of \emph{degradable} channels~\cite{degradable,Structure,AdditiveExtension} are strongly additive when combined with another degradable channel, and qubit unital channel~\cite{Additivity}, entanglement-breaking channels~\cite{Shor2002} as well as depolarizing channels~\cite{King2003} have strongly additive Holevo capacity and classical capacity.

Recently, there is a quantum channel with input-output dimensions $3$ and the environment dimension $2$ called the qutrit platypus channels $\N_s$~\cite{Lovasz,Singular,platypusTIT} attracting attention, which relies on a two value probability $(s,1-s)$ defined in $\mathbb{R}^2$. In particular, for the special case $s = 1/2$, it can be generalized to the channel $\M_{d+1}$ with input-output dimension $d+1$ and environment dimension $d$, which corresponds to a $d$-dimensional discrete uniform distribution. Such channels have easily computable coherent information, and the quantum capacity of $\N_s$ and $\M_{d+1}$ are both controlled by the maximum value of their corresponding probability distributions~\cite{SDPBoundC}. Therefore, the quantum capacity of $\M_{d+1}$ can become arbitrarily small when the dimension $d$ increases. As only two types of quantum channels are known with vanishing quantum capacity, anti-degradable channels and positive partial transpose channels, $\M_{d+1}$ provides a new understanding of the channel structure for quantum capacity small enough. Moreover, $\N_s$ and $\M_{d+1}$ has weakly additive private and classical capacities, and they do not belong to the previously known class of channels with weak additivity that we mentioned above. Further, when combines $\M_{d+1}$ with the qudit erasure channel $\mathcal{E}_{\lambda,d}$, the quantum capacity is super-additive~\cite{platypusPRL}, thus showing a special phenomenon called ``near-super-activation'' which is close the famous ``super-activation''~\cite{superactivation,Superact2}, where super-activation combines two channels with zero quantum capacity to achieve a positive quantum communication rate and near-super-activation can also achieve a positive quantum capacity by combining a channel with zero and another with arbitrarily small quantum capacity.

In this work, we consider a more general form of platypus channels $\mathcal{O}_{\vec{\mu}}$ defined with respect to an arbitrary $d$-dimensional probability distribution $\vec{\mu}$ which has input-output dimensions $d+1$ and environment dimension $d$. This channel $\calO$ has simple coherent information and its quantum capacity is also controlled by the maximum probability value of $\vec{\mu}$ which completely generalizes the previous results about $\N_s$ and $\M_{d+1}$ into any probability distribution. In addition, the private and classical capacity of $\calO$ are both weakly additive and equal to $1$ for any probability vector $\vec{\mu}$, showing a wide class of quantum channels that do not belong to any known class of quantum channels with positive gap between quantum and private capacities or weak additivity of private and classical capacities. Furthermore, when combining $\mathcal{O}_{\vec{\mu}}$ with two different channels with zero quantum capacity - qudit erasure channels $\mathcal{E}_{\lambda,d}$ and multilevel amplitude damping channels $\mathcal{A}_{\gamma}$ respectively, they both exhibit the super-additivity of quantum capacity. Therefore, as long as the proper probability distribution $\vec{\mu}$ is chosen to generate the generalized platypus channel $\calO$ with arbitrarily small quantum capacity, near-super-activation is widely available.

Our work generalizes the results in previous work~\cite{platypusTIT,platypusPRL}. While prior works have utilized the qudit erasure channel, we also provide a fresh quantum channel that can also assist in achieving super-additivity of quantum capacity. We hope our approach will bring new perspectives to the subject.

\section{Results}
\subsection{The model of generalized platypus channels}

Let $A$ and $B$ denote finite-dimensional quantum systems with Hilbert spaces $\mathcal{H}_A$ and $\mathcal{H}_B$, respectively. A \emph{quantum channel} from $A$ to $B$ consists of a completely positive, trace-preserving (CPTP) linear map 
$
\mathcal{N}:\mathcal{L}(\mathcal{H}_A)\longrightarrow \mathcal{L}(\mathcal{H}_B)\, ,
$
where $\mathcal{L}(\mathcal{H})$ denotes the algebra of linear operators on a Hilbert space $\mathcal{H}$. Given a quantum channel $\N$, there exists an environment $E$ and an isometric embedding $V:\mathcal{H}_A\longrightarrow \mathcal{H}_B\otimes \mathcal{H}_E$ such that $\N(\rho) = \tr_E(V \rho V^\dagger)$ for all $\rho\in \mathcal{L}(\mathcal{H}_A)$, which is referred to as a \emph{Stinespring representation} of the channel $\mathcal{N}$. The complementary channel to $\N$ is the channel $\N^c:\mathcal{L}(\mathcal{H}_A)\to \mathcal{L}(\mathcal{H}_E)$ obtained by tracing out over $A$, i.e., $ \N^c(\rho) = \tr_A(V \rho V^\dagger)$. A channel $\mathcal{N}$ is called \emph{degradable} if there is a degrade channel $\mathcal{W}$ such that $\mathcal{W}\circ \mathcal{N} = \mathcal{N}^c$, while the channel $\N$ is called \emph{anti-degradable} if it complement $\N^c$ is degradable.

The coherent information of $\N$ is given by the formula
\begin{equation}\label{Q1}
    \Q(\N)=\max_{\rho}I_c(\rho,\N)\, ,
\end{equation}
where $I_c(\rho,\N)$ is
\[
I_c(\rho,\mathcal{N})=S(\N(\rho))-S(\N^c(\rho))\, ,
\]
and $S(\cdot)$ denotes von~Neumann entropy. As we mention above, the degradable channel $\N$ has weakly additive coherent information, thus $Q(\N) = \Q(\N)$. Moreover, due to no-clone theorem, the quantum capacity of an anti-degradable channel always vanishes.

Let $A$ is the input system with dimension $d+1$ and $\{|i\rangle\}_{i=0}^d$ is a computational basis of $A$. Given a probability distribution $\vec{\mu} = (\mu_0,\cdots,\mu_{d-1})$, the generalized platypus channel $\mathcal{O}_{\vec{\mu}}$ is defined by the isometry $V: A\longrightarrow BE$:
\begin{equation}\label{unitary}
    \begin{aligned}
        V |0\rangle & = \sum_{j=0}^{d-1} \sqrt{\mu_j}\, |j\rangle\otimes|j\rangle \,, \\
        V |i\rangle & = |d\rangle\otimes|i-1\rangle\,,\ \text{for}\ i = 1,\cdots, d\,,
    \end{aligned}
\end{equation}
where the output system have dimension $d+1$, the environment is with dimension $d$, and $\mathcal{O}_{\vec{\mu}}(\cdot) = \tr_E (V\cdot V^\dagger)$. Without loss of generalized, we can assume that $\mu_0\leq\cdots\leq \mu_{d-1}$ since when $\mu_0 > \mu_1$, we can exchange $|0\rangle$ with $|1\rangle$ so that the isometry is defined by the probability distribution with increasing entries. 

\subsection{Capacities of the generalized platypus channel}

The generalized platypus channel is firstly introduced in~\cite{platypusTIT} where $\calO$ is proved that its coherent information $\Q(\mathcal{O}_{\vec{\mu}})$ is attained on the state in the form of $\rho(u) = (1-u)|0\rangle\langle 0| + u|d\rangle\langle d|$,
\begin{equation}\label{Q1ofO}
    \Q(\calO) = \max_{u\in[0,1]} I_c\big(\rho(u),\calO\big) > 0\,.
\end{equation}
One can also show the positivity of its coherent information by noticing the output states of $\calO$ with respect to the input state $|0\rangle\langle 0|$ and $|d\rangle\langle d|$ is orthogonal, but the output states of $\calO^c$ is not, together with the criterion in~\cite{DetectingPositive_NPJ}. Moreover, assuming the validity of spin-alignment conjecture~\cite{platypusTIT,SpinAlign}, the coherent information of $\calO$ is weakly additive so that $Q(\calO) = \Q(\calO)$. Without the conjecture, according to the well-known ``transposition bound''~\cite{TranspositionBound}: $Q(\B) \leq \log ||\T\circ \B||_\diamond$ for a generic channel $\B$ where $\T$ is the transpose map and $||\cdot||_\diamond$ is the diamond norm which can be solved by SDP, we prove that the quantum capacity of $\calO$ has an upper bound relying on the maximum probability value of $\vec{\mu}$.
\begin{thm}\label{thm1}
   Given a probability distribution $\vec{\mu}$, the quantum capacity of $\calO$ is bounded from above as
    \begin{equation}\label{Qbound}
        Q(\calO) \leq \log (1+\sqrt{\max_i \mu_i})\,.
\end{equation}
\end{thm}
This bound unifies and generalizes previous results on upper bounds on the quantum capacity of $\mathcal{N}_s$ and $\mathcal{M}_d$~\cite{platypusTIT}, and indicates that the quantum capacity of $\calO$ is completely controlled by the largest probability value. In particular, as the dimension $d$ increases, it is possible to choose a probability distribution whose maximum probability value small enough so that the quantum capacity of the corresponding generalized platypus channel $\calO$ is also sufficiently close to zero.
Moreover, based on this result, it is natural to ask whether the maximum probability value also determines the strong super-additivity of quantum capacity, which is confirmed by our later examples. Although the transposition bound is improved by other SDP upper bounds~\cite{SDPBoundC,Fang2021}, we still use it to obtain an upper bound of quantum capacity as Eq.~\eqref{Qbound} for its analytical expression and dependence on the maximum probability value. 

While the quantum capacity portrays the maximum rate for asymptotically transmitting quantum information with vanishing errors using the quantum channel $\B$, the private $P$ and classical $C$ capacity describes the maximum rate for the quantum channel $\B$ asymptotically transmitting private and classical information with vanishing errors respectively, which are given by
\begin{equation*}
    P(\B) = \lim_{n\rightarrow \infty} \frac{P^{(1)}(\B^{\otimes n})}{n}\,, \, C(\B) = \lim_{n\rightarrow \infty} \frac{C^{(1)}(\B^{\otimes n})}{n}
\end{equation*}
where the private information $P^{(1)}$ and Holevo capacity $C^{(1)}$ are defined by taking the maximum of $I_p$ and Holevo information $\chi$ over the ensemble $\{p_i,\rho_i\}$ respectively, that is
\begin{equation*}
    P^{(1)} = \max_{\{p_i,\rho_i\}} I_p\Big(\{p_i,\rho_i\}, \B\Big)\,, \,  C^{(1)} = \max_{\{p_i,\rho_i\}} \chi\Big(\{p_i,\rho_i\}, \B\Big)
\end{equation*}
where $I_p$ and Holevo information $\chi$ of the channel $\B$ with respect to an ensemble $\{p_i,\rho_i\}$ is defined as 
\begin{equation*}
    \begin{aligned}
        I_p\Big(\{p_i,\rho_i\}, \B\Big) & = I_c\Big(\sum_i p_i\rho_i\,, \B\Big) - \sum_i p_i I_c(\rho_i,\B)\,, \\
        \chi\Big(\{p_i,\rho_i\}, \B\Big) & = S\Big(\B\big(\sum_i p_i\rho_i\big)\Big) - \sum_i p_i S\big(\B(\rho_i)\big)\,.
    \end{aligned}
\end{equation*}
On the other hand, the entanglement-assist classical capacity $C_E$~\cite{ECC} measures the maximum rate for a quantum channel $\mathcal{B}$ transmitting classical information with the assistance of unlimited prior entanglement between the sender and receiver, which is always additive:
\begin{equation*}
    C_E(\mathcal{B}) = \max_{\rho} S(\rho) + I_c(\rho,\mathcal{B}) = \max_\rho I(\rho,\mathcal{B}) \,,
\end{equation*}
where $I(\rho,\mathcal{B})$ is the mutual information of the quantum channel $\mathcal{B}$ with the input state $\rho$. Using the SDP bound for classical capacity~\cite{SDPBoundC}, we can evaluate the exact value of private and classical capacity for the generalized platypus channel.  
\begin{thm}\label{thm2}
    Let $\calO$ is defined above associated with a probability distribution $\vec{\mu}$, then the private and classical capacity and the entanglement-assist classical capacity of $\calO$ are 
    \begin{equation}
    \begin{aligned}
        P^{(1)}(\calO) = C^{(1)}(\calO) & = P(\calO) = C(\calO) = 1 \,, \\
        C_E (\calO) & = 2\,.
    \end{aligned}
    \end{equation}
\end{thm}

\begin{figure*}
    \centering
    \subfigure[d = 10]{\includegraphics[width=\columnwidth]{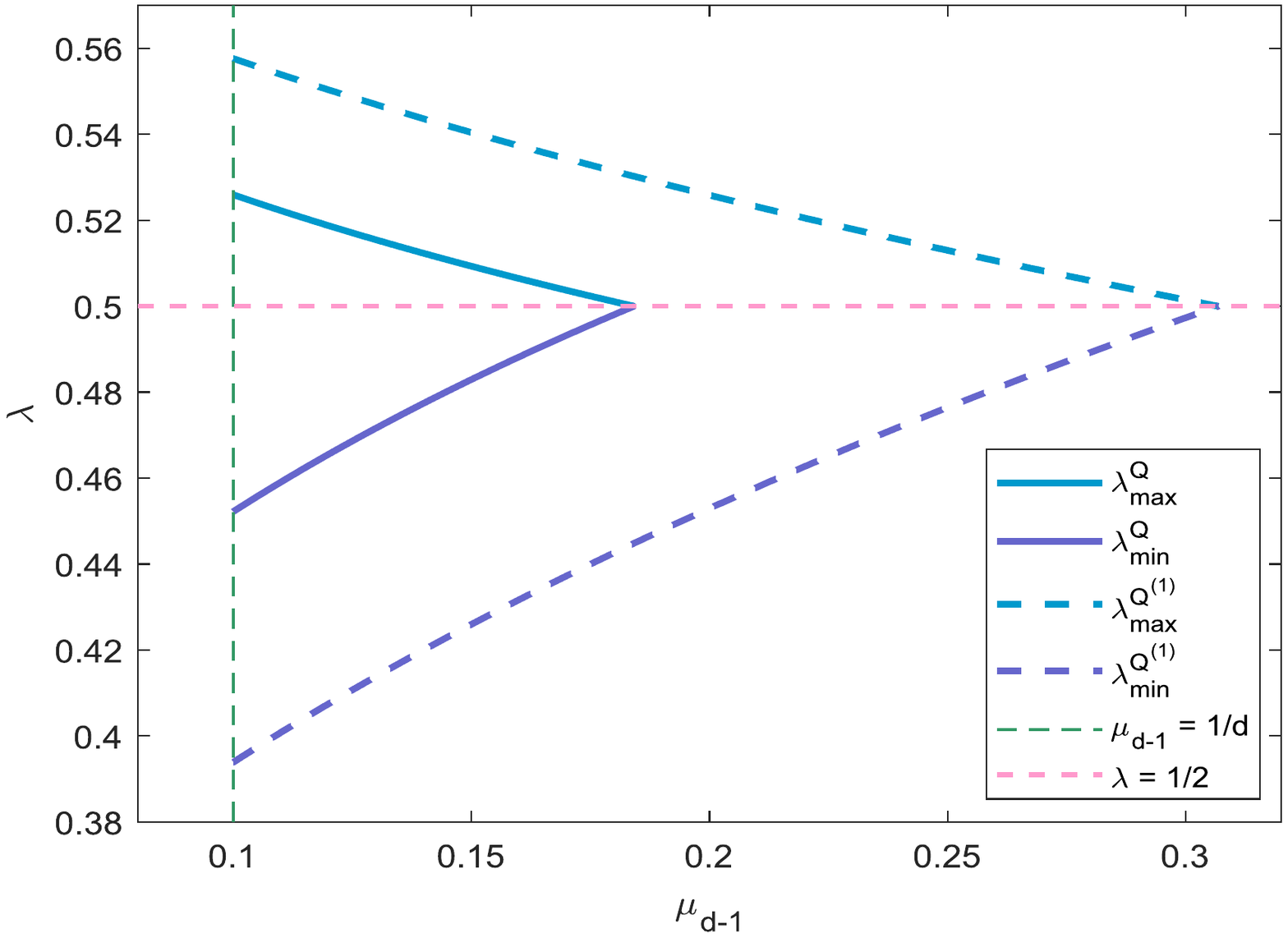}}
    \subfigure[d = 50]{\includegraphics[width=\columnwidth]{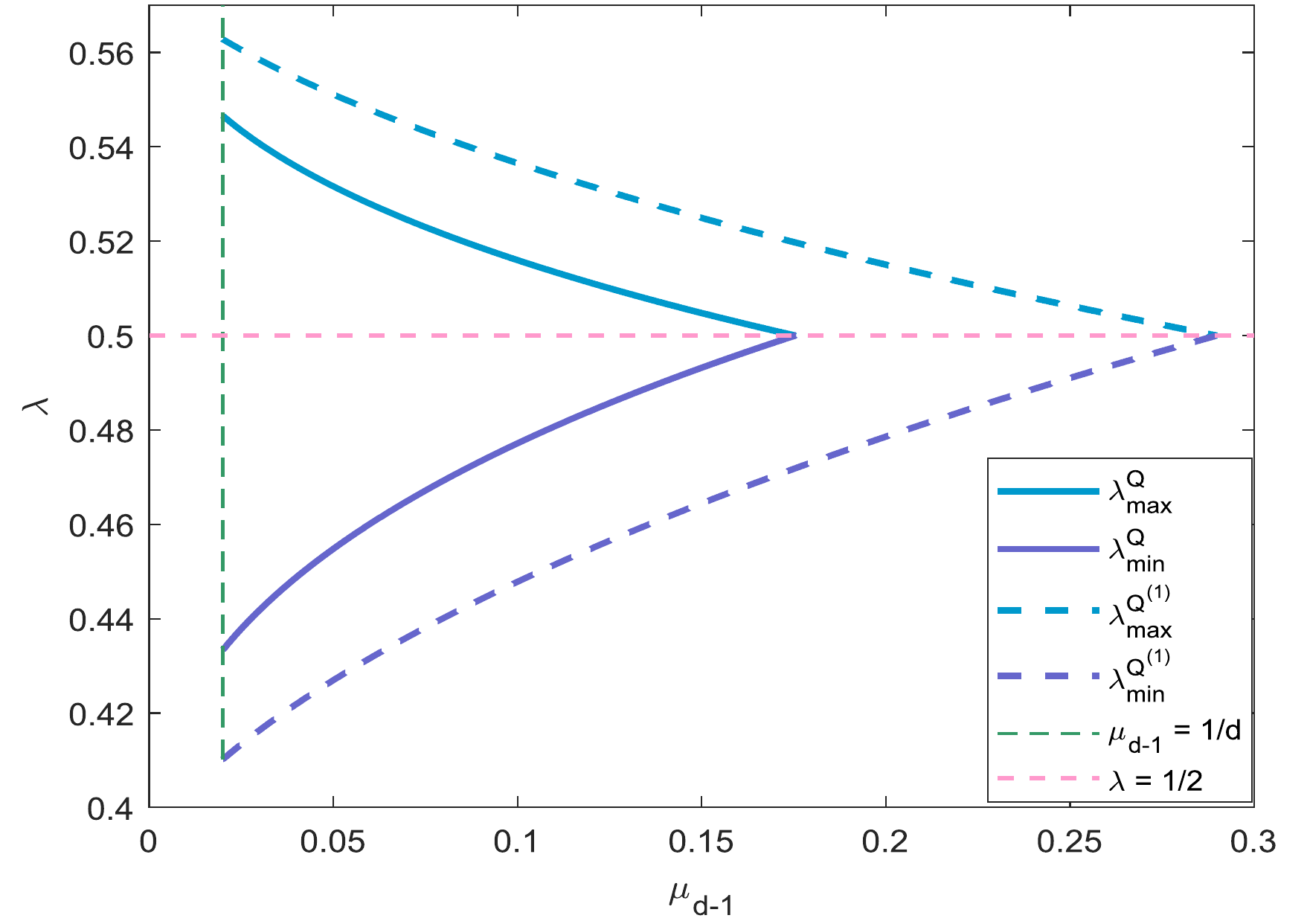}}
    \caption{The range of $\lambda$ and $\mu_{d-1}$ for super-additivity of quantum capacity for $\calO$ and $\mathcal{E}_{\lambda,d}$. The smallest $\mu_{d-1}$ is $1/d$ for different $d$ which is the green dashed line. The  blue solid line on the top is the maximum $\lambda_{\max}^{Q}$, while the purple solid line below is the minimum $\lambda_{\min}^{Q}$ with respect to $\mu_{d-1}$ such that $\calO$ has super-additive quantum capacity $Q$. The blue dashed line on the top is the maximum $\lambda_{\max}^{\Q}$, while the purple dashed line below is the minimum $\lambda_{\min}^{\Q}$ with respect to $\mu_{d-1}$ such that $\calO$ has super-additive coherent information $\Q$.}
    \label{fig1}
\end{figure*}

Since the quantum information is necessary private which in turn requires classical communication, a natural inequality holds:
\begin{equation*}
    Q(\B) \leq P(\B) \leq C(\B)\,.
\end{equation*}
Generally, it's also hard to determine the private and classical capacity, hence the strict gap between these capacities is also unknown except few special classes of channels like (regularized) less noisy, more capable channels~\cite{Watanabe,Hirche2022,IEEEPartialOrder}, the Horodecki channel~\cite{Horodeckichannel1,HorodeckiChannel2,Horodeckichannel3,Horodeckichannel4} and the `half-rocket' channel~\cite{Leung2014}. Here the generalized platypus channel has nice channel capacities shown above and according to Eq.~\eqref{Qbound}, it also exhibits the positive gap between quantum and private capacity except the probability vector $\vec{\mu}$ is trivial as $(0,\cdots,0,1)$. It is worth mentioning that generalized platypus channels do not belong to any previously known class of channels with additive private and classical capacity, which may give us some new understanding of channel capacity theory.

\subsection{Super-additivity of quantum capacity of \texorpdfstring{$\calO$}{} combined with the qudit Erasure channel}
The qudit erasure channel $\mathcal{E}_{\lambda,d}$ keeps the input state with probability $1-\lambda$ and replace it  with probability $\lambda$ as a pure state $|e\rangle\langle e|$ orthogonal to all the input states. Formally, let $A'$ is an input system with dimension $d$ and $B'$ is the output system with dimension $d+1$, the erasure channel acts on the state of $A'$ as:
\begin{equation*}
    \mathcal{E}_{\lambda,d}(\rho) = (1-\lambda)\,\rho + \lambda\, |e\rangle\langle e|\,.
\end{equation*}
Its complementary channel is also an erasure channel with erasure probability $1-\lambda$, that is $\mathcal{E}_{\lambda,d}^c = \mathcal{E}_{1-\lambda,d}$, hence one can find a degrade channel $\mathcal{W}$ such that $\mathcal{W}\circ \mathcal{E}_{\lambda,d} = \mathcal{E}_{\lambda,d}^c$ for $\lambda \in [0,1/2]$, which implies the qudit erasure channel $\mathcal{E}_{\lambda,d}$ is degradable for $\lambda \in [0,1/2]$ and anti-degradable for $\lambda \in [1/2,1]$. So the quantum capacity of the qudit erasure channel is~\cite{Erasurechannel} 
\begin{equation}\label{QofE}
    Q(\mathcal{E}_{\lambda,d}) = \max \big\{(1-2\lambda)\log d,0\big\} \,.
\end{equation}

The Ref.~\cite{platypusPRL} shows that the channel $\M_{d+1}$ corresponding to the uniform probability distribution $\vec{\mu} = (\frac{1}{d},\cdots,\frac{1}{d})$ demonstrates the super-additivity of quantum capacity when combined with the qudit erasure channel $\mathcal{E}_{\lambda,d}$. Here we prove that such super-additivity can exist for a larger range of $\vec{\mu}$. More precisely, combining the qudit erasure channel $\mathcal{E}_{\lambda,d}$ and the generalized platypus channel $\calO$ associated with a probability distribution $\vec{\mu}$ with dimension $d$, we show that 
\begin{thm}\label{thmE}
    Given a probability distribution $\vec{\mu}$ corresponding to the generalized platypus channel $\calO$, without loss of generality, suppose $\max_i \mu_i = \mu_{d-1} \in [1/d,3-2\sqrt{2}]$ for $d$ large enough, then there exists $\lambda$ such that the quantum capacity of $\calO$ and $\mathcal{E}_{\lambda,d}$ is super-additive, that is
\begin{equation}\label{SuperAddQ}
    Q(\calO\otimes \mathcal{E}_{\lambda,d}) > Q(\calO) + Q(\mathcal{E}_{\lambda,d})\,.
\end{equation}
\end{thm}
It's clear the condition holds for $d\geq 6$, that is one can always find a generalized platypus channel defined on high dimensions with super-additive quantum capacity. Particularly, solving the eigenvalue problem~\eqref{B1} in Appendix, we can get a tighter result: the super-additivity of quantum capacity of $\calO$ exists for $d \geq 4$, for instance when $d=4$, $\M_5$ is an example as shown in~\cite{platypusPRL}. When consider the super-additivity of coherent information $\Q$, the condition on the maximum value of $\vec{\mu}$ can be attenuated to $\max_i \mu_i < 0.28$. Again, solving the eigenvalue problem can give a tighter result as the super-additivity of $\Q$ of $\calO$ exists for $d \geq 3$. 

As shown in Fig.~\ref{fig1}, the range of $\mu_{d-1}$ and $\lambda$ so that $\calO$ has super-additive quantum capacity and coherent information when combined with $\mathcal{E}_{\lambda,d}$, are in the similar shape for various $d$. when $\mu_{d-1} = 1/d$ is fixed, the relation between $\lambda$ and $d$ are obtained in~\cite{platypusPRL}. While fixed $\lambda = 1/2$, it gives the maximum $\mu_{d-1}$ keeping the super-additivity of both quantum capacity and coherent information. Moreover, this maximum $\mu_{d-1}$ decreases with respect to $d$, but is always great than $3-2\sqrt{2}$ for super-additivity of $Q$ which as claimed in the Theorem~\ref{thmE}; and the maximum $\mu_{d-1}$ is always bigger than $0.282$ for super-additivity of coherent information $\Q$ for $\calO$ and the 50-\% erasure channel in any dimension $d \geq 3$.

In order to prove these results, suppose $A'$ is a $d$-dimension Hilbert space with a computational basis $\{|i\rangle\}_{i=0}^{d-1}$, we consider the bipartite state $\rho$ as
\begin{equation}\label{state}
    \rho = \frac{1}{2} |0\rangle\langle0|_A\otimes \frac{\mathds{1}_{A'}}{d} + \frac{1}{2} |\psi\rangle\langle \psi|_{AA'}\,,
\end{equation}
where $|\psi\rangle_{AA'} = \sum_{i=1}^d \sqrt{\mu_{i-1}} |i\rangle_A\otimes|i-1\rangle_{A'}$ is a bipartite entangled state. Evaluating the output entropy of $\calO\otimes \mathcal{E}_{\lambda,d}$ and its complement with respect to this input state, and using the Weyl inequality~\cite{Weylinequality,MatrixAnalysisHorn} and weak majorization, we obtain
\begin{equation}
    I_c\Big(\rho, \calO\otimes \mathcal{E}_{\lambda,d}\Big) \geq 1-\lambda + \frac{d - (2d-\mu_{d-1})\lambda}{2d} \log d\,.
\end{equation}
Then utilizing the upper bound of $Q(\calO)$ presented in Eq.~\eqref{Qbound} and the exact solution coherent information in Eq.~\eqref{Q1ofO}, together with the quantum capacity of $\mathcal{E}_{\lambda,d}$ in Eq.~\eqref{QofE}, we arrive at the result about super-additivity of quantum capacity. 

\subsection{Super-additivity of quantum capacity of \texorpdfstring{$\calO$}{} combined with the multilevel amplitude damping channel}

Combined with qudit erasure channels gives rise to super-additivity of quantum capacity which can be regarded as ``near-super-activation''~\cite{platypusPRL} since the quantum capacity of the generalized platypus channel $\calO$ with high dimension can vanish. We now offer a fresh quantum channel that also exhibits super-additivity of quantum capacity for $\calO$.

\begin{figure}
    \centering
    \includegraphics[width=\columnwidth]{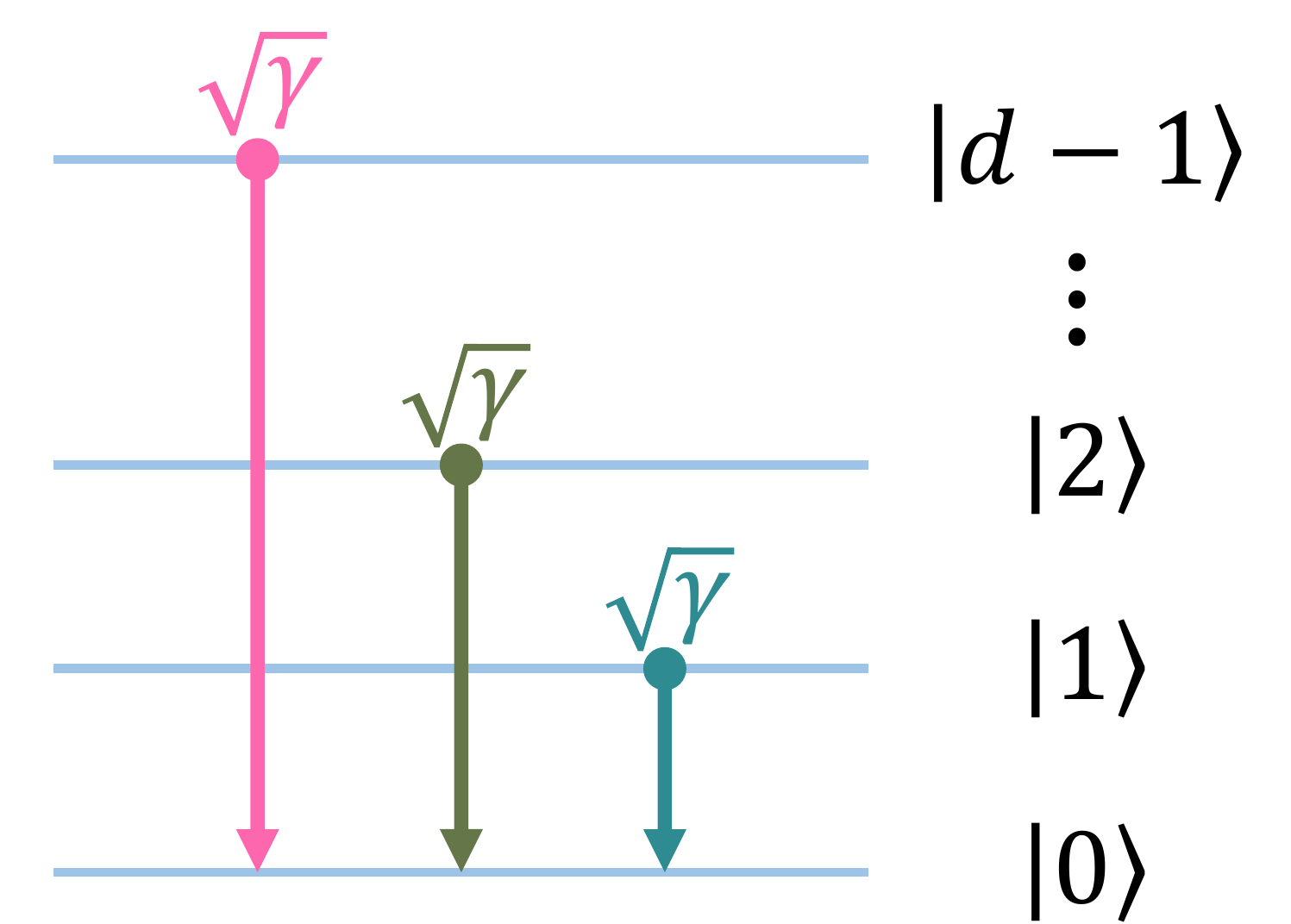}
    \caption{Schematic representation of the multilevel amplitude damping channel $\mathcal{A}_{\gamma}$ we considered. It acts on the system with dimension $d$, every arrow represents a decaying process from $j$ to $0$ at a rate $\gamma$ where the ground state $|0\rangle$ is fixed.}
    \label{fig2}
\end{figure}

Suppose $A'$ is a $d$-dimension Hilbert space with a computational basis $\{|i\rangle\}_{i=0}^{d-1}$, the multilevel amplitude damping channel $\mathcal{A}_\gamma$ considered here is schematized in Fig.~\ref{fig2} whose Kraus operators $\{K_i\}_{i=0}^{d-1}$ are
\begin{equation}
\begin{aligned}
    K_0 & = |0\rangle\langle 0| + \sqrt{1-\gamma} \sum_{j=1}^{d-1} |j\rangle\langle j|\,,\\
    K_j & = \sqrt{\gamma} |0\rangle\langle j|\,, \quad \text{for}\quad j = 1,\cdots,d-1\,.
\end{aligned}
\end{equation}
Its complementary channel $\mathcal{A}^c_\gamma$ is also an amplitude damping channel with decaying rate $1-\gamma$. Since $\mathcal{A}_{\frac{1-2\gamma}{1-\gamma}}\circ \mathcal{A}_{\gamma} = \mathcal{A}_{\gamma}^c$, one can obtain $\mathcal{D}_\gamma$ is degradable for $\gamma\in [0,1/2]$ and anti-degradable for $\gamma\in [1/2,1]$. Moreover, as $\mathcal{A}_{\gamma}$ is covariant under the group containing the unitary transformations which are diagonal in the computational basis, its quantum capacity has single-letter expression and is obtained on the diagonal state $\sigma_{\text{diag}}$~\cite{GADC1,GADC2}, calculate the exact expression of coherent information, we can get the quantum capacity of $\mathcal{A}_{\gamma}$ is given by
\begin{equation*}
    Q(\mathcal{A}_\gamma) = \Q(\mathcal{A}_\gamma) = \max_{\sigma(u)}\, I_c\big(\sigma(u),\mathcal{A}_\gamma\big)
\end{equation*}
where $\sigma(u) = \text{diag}(u,\frac{1-u}{d-1},\cdots,\frac{1-u}{d-1})$ and $\text{diag}(\vec{\alpha})$ is the diagonal matrix with the entry of $\vec{\alpha}$ in its diagonal part.

\begin{figure*}
    \centering
    \subfigure[d = 10]{\includegraphics[width=0.32\textwidth]{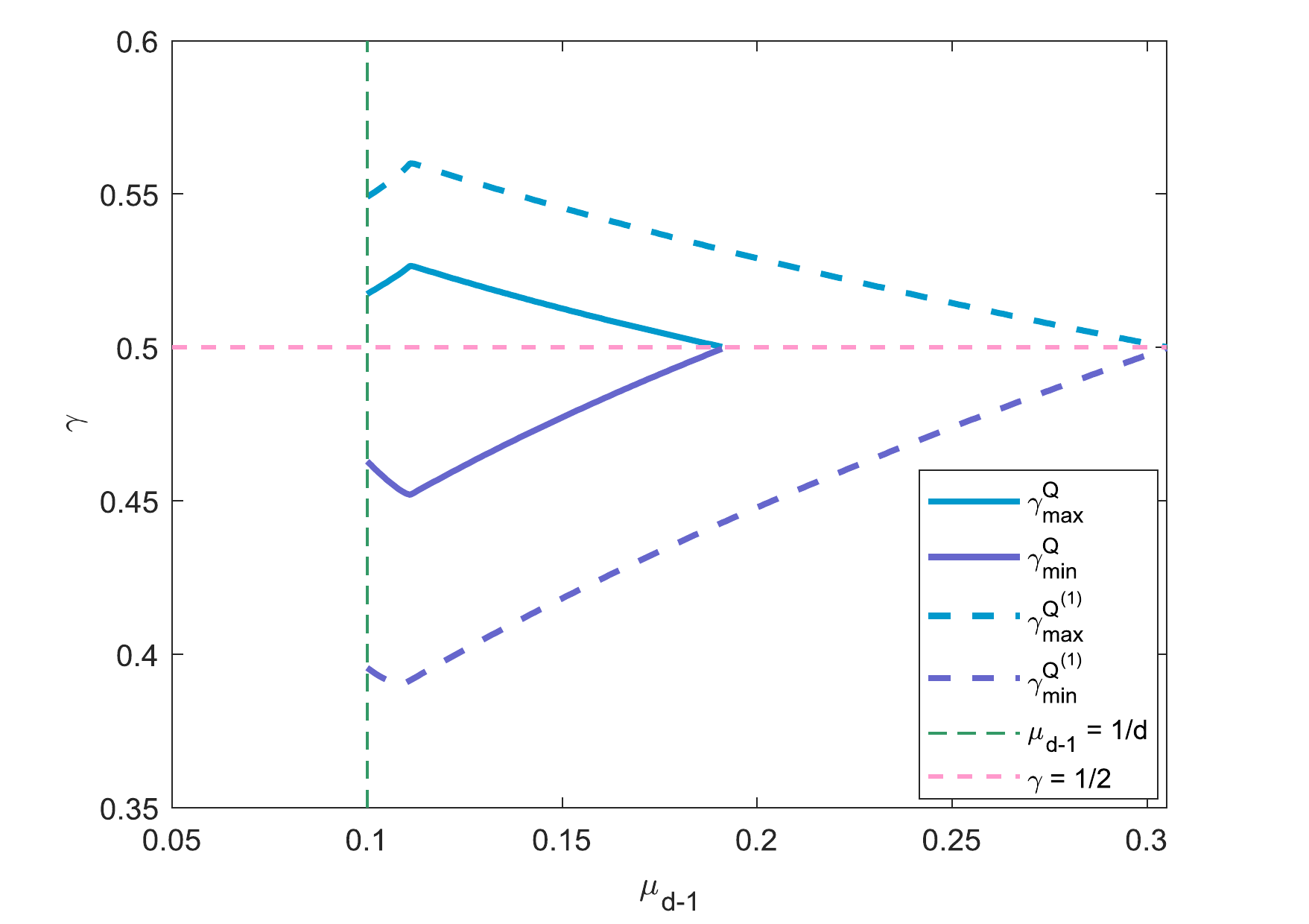}}
    \subfigure[d = 50]{\includegraphics[width=0.32\textwidth]{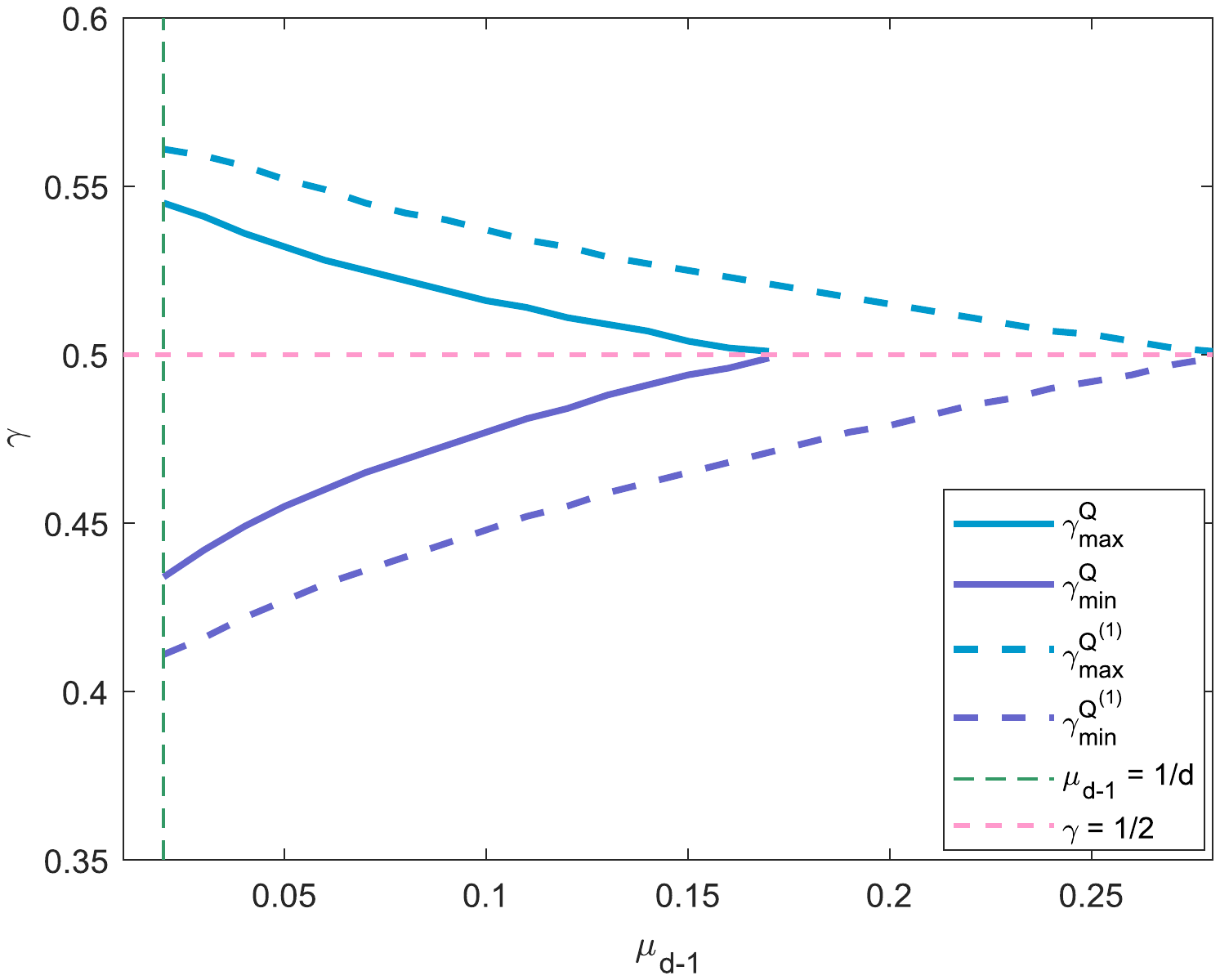}}
    \subfigure[$\mu_i = 1/d$]{\includegraphics[width=0.32\textwidth]{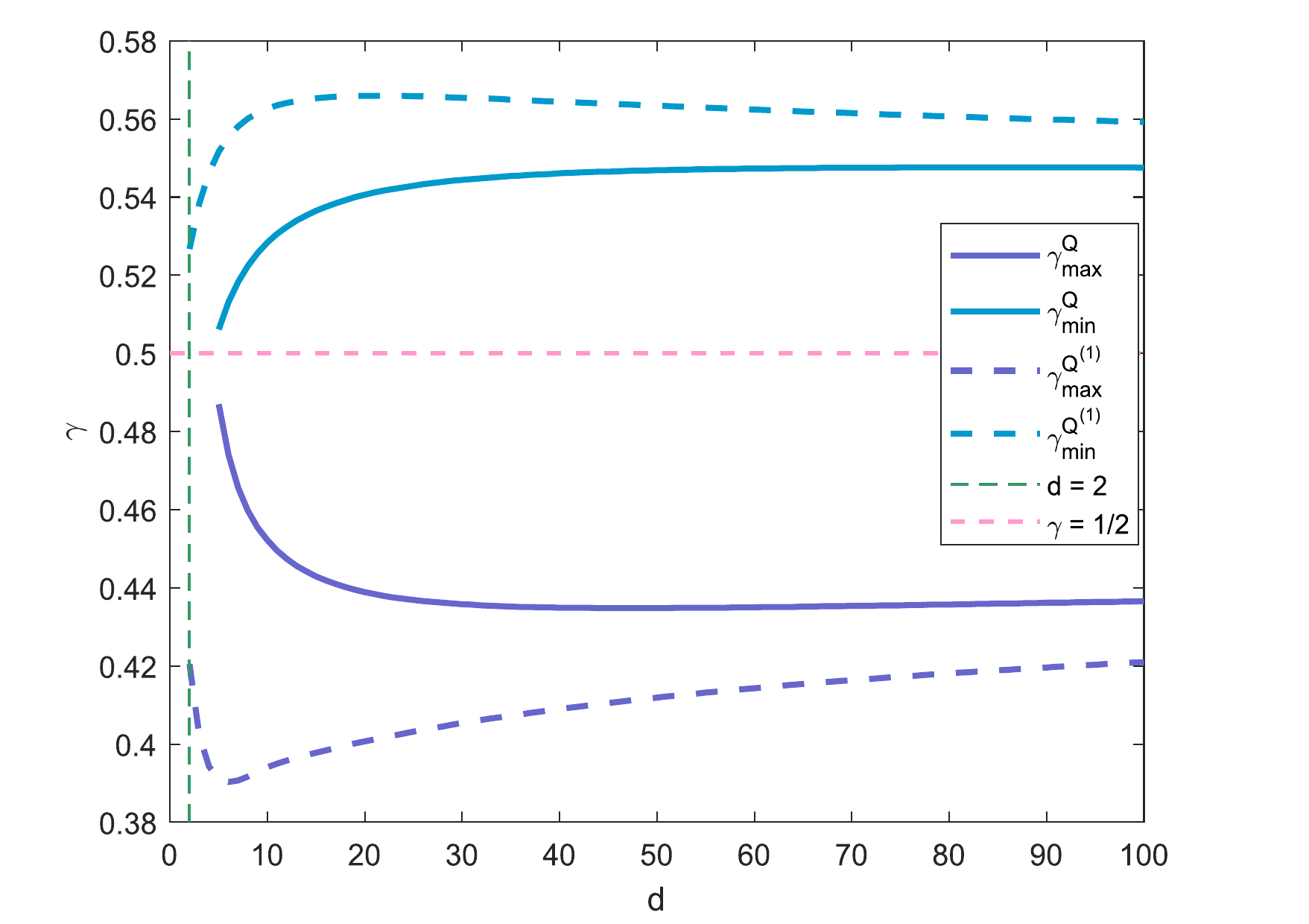}}
    \caption{The range of $\gamma$ and $\mu_{d-1}$ for super-additivity of quantum capacity for $\calO$ and $\mathcal{A}_{\gamma}$. (a) and (b): the x-axis is the maximum value $\mu_{d-1}$ in the probability vector $\vec{\mu}$ and the y-axis is about $\gamma$ so that $\calO$ has super-additive quantum capacity and coherent information combined with $\mathcal{A}_{\gamma}$. (c): Set $\vec{\mu} = \frac{1}{d}(1,\cdots,1)$, the range of $\gamma$ for $\M_{d+1}$ with super-additivity of quantum capacity and coherent information. Both the blue lines on the top (dashed or solid) is about the maximum $\gamma$ while the purple lines below is about the minimum $\gamma$ for the super-additivity of $Q$ and coherent information $\Q$.}
    \label{fig3}
\end{figure*}

Inputting the state $\rho$ in Eq.~\eqref{state} to $\calO\otimes\mathcal{A}_{\gamma}$ and its complement, the exact value of $I_c(\rho,\calO\otimes\mathcal{A}_\gamma)$ is hard to give an analytical result but can be bounded below which only relies on the maximum probability value $\mu_{d-1}$ and $\gamma$ for fixed $d$ as discussed in Appendix. We display the range of $\mu_{d-1}$ and $\gamma$ for $d = 10$ and $ d = 50$ such that $\calO$ has super-additive quantum capacity and coherent information when combined with $\mathcal{A}_{\gamma}$ in Fig.~\ref{fig3}~(a),(b). Next, let's discuss some special cases. 

First, when $\gamma = 1/2$ is fixed, the multilevel amplitude damping channel is self-complementary~\cite{selfcomplementary}, that is $\mathcal{A}_{1/2} = \mathcal{A}_{1/2}^c$, hence its quantum capacity vanishes. Since the coherent information has lower bound:
\begin{equation}
    I_c\big(\rho,\calO\otimes\mathcal{A}_{1/2}\big) \geq \frac{1-\mu_{0}}{2}\,,
\end{equation}
where $\mu_0$ is the minimum value of $\vec{\mu}$, using the trivial inequalities $\log(1+\sqrt{\mu_{d-1}}) \leq \sqrt{\mu_{d-1}}\leq \sqrt{1-(d-1)\mu_{0}}$, then for $d\geq 6$, there exists a probability vector $\vec{\mu}$ such that the generalized platypus channel $\calO$ has super-additive quantum capacity when combined with $\mathcal{A}_{1/2}$.

Second, let the probability vector $\vec{\mu} = \frac{1}{d}(1,\cdots,1)$, i.e., $\calO = \mathcal{M}_{d+1}$,
we can compute the exact value of $I_c(\rho,\M_{d+1}\otimes\mathcal{A}_\gamma)$, which only depends on $\gamma$, and we show the range of $\gamma$ such that $\M_{d+1}$ has super-additive quantum capacity and coherent information when combined with $\mathcal{A}_{\gamma}$ in Fig.~\ref{fig3}(c). The minimum dimension for the super-additivity of quantum capacity for $\calO$ and $\mathcal{A}_{\gamma}$ is $d_{\min}^{Q} = 5$, while $d_{\min}^{\Q} = 2$ is the minimum dimension so that $\M_{d+1}$ and $\mathcal{A}_{\gamma}$ has super-additive coherent information as shown in~\cite{platypusPRL}. More important, as $d$ increases, the region about $\gamma$ required by super-additive quantum capacity of $\calO$ and $\mathcal{A}_{\gamma}$ is converging to the range of $\gamma$ for super-additive coherent information $\Q$. This indicates that the super-additivity of quantum capacity for the generalized platypus channel $\calO$ exhibits a similar behavior to the super-additivity of $\Q$, which provides new examples and perspectives for the subsequent study of quantum Shannon theory.

\section{Discussion}

In this paper, we consider the generalized platypus channel $\calO$, with input and output dimensions $d+1$ and environment dimension $d$, defined corresponding to a probability vector $\vec{\mu}$ and the unitary in Eq.~\eqref{unitary}, and compute its quantum channel capacities such as quantum, private and classical capacity. Moreover, when combined with qudit erasure channels and multilevel amplitude damping channels respectively, the quantum capacity of the generalized platypus channel is super-additive and such increment can be as big as $1/2$.

The quantum capacity of the generalized platypus channel $\calO$ is controlled by the maximum probability $\max_i \{\mu_i\}$ in the probability vector $\vec{\mu}$ as Eq.~\eqref{Qbound}
$$
Q(\calO) \leq \log (1+\sqrt{\max_i \mu_i})\,.
$$
Since the maximum value of the probability vector can tends to zero as the dimension $d$ increases, this upper bound can likewise converge to zero. Therefore, the generalized platypus channel is an important example whose quantum capacity can be arbitrarily small. Moreover, taking advantage of the SDP bound for classical capacity of a generic quantum channel, the private information and Holevo capacity of $\calO$ are proven to be both weakly additive and hence its private and classical capacities equal to $1$. The previously known channels with weakly additive private information $P^{(1)}$ are less noisy channels~\cite{Watanabe,IEEEPartialOrder} which satisfy $P(\mathcal{B}^c) = 0$, anti-degradable channels~\cite{LSD2} and direct sum channels~\cite{DirectSumC}. Whereas channel capacities of the complement for generalized platypus channels are $Q(\calO^c) = P(\calO^c) = C(\calO^c) = \log d$~\cite{platypusTIT}, and thus $\calO$ does not belong to these classes of channels mentioned above. It is worthwhile to further investigate the structure satisfied by the generalized platypus channel to obtain a new and broader class of quantum channels having weak additive one-shot private capacity.

Except the trivial probable vector $\vec{\mu}$, there always is a positive gap between the quantum and private capacity of $\calO$, which provides a new example for the separation of these capacities different from the Horodecki channel~\cite{Horodeckichannel1,HorodeckiChannel2,Horodeckichannel3,Horodeckichannel4} and half-rocket channels~\cite{Leung2014}. Since Horodecki channels play an important role in quantum superactivation and the half-rocket channel exhibits the super-additivity of private capacity, while we proved that the generalized platypus channel has super-additive quantum capacity, thus we hope that subsequent studies to show whether $\calO$ has super-additive private or classical capacity.

In previous results showing super-additivity of quantum or private capacities, they often combines special channels with qudit 50-\% erasure channel. In contrast, our results indicate that we can slightly relax the condition on the erasure probability of erasure channels, i.e., no longer require that the erasure probability be 50-\%, but rather an interval. This conforms our intuition since the quantum channel capacity is continuous~\cite{ContinuityQPC}. Furthermore, we also provide a new channel: the multilevel amplitude damping channel $\mathcal{A}_{\gamma}$, that has properties similar to qudit erasure channels, such as degradability and anti-degradability. In particular, they are both self-complementary for a special parameter. When combined with the generalized platypus channel, the multilevel amplitude damping channel is also capable of achieving super-additivity of quantum capacity. Hence, we would like to wonder whether $\mathcal{A}_{\gamma}$, or more generally, self-complementary channels, can replace the role of the qudit 50-\% erasure channel in previous work about super-additivity of channel capacities.

\section{Acknowledgments} 

Z.M. is supported by the Fundamental Research Funds for the Central Universities, the National Natural Science Foundation of
China (no. 12371132). Q.Z. acknowledges funding from the National Natural Science Foundation of China (NSFC) via Project No. 12347104 and No. 12305030, Guangdong Natural Science Fund via Project 2023A1515012185, Hong Kong Research Grant Council (RGC) via No. 27300823, N\_HKU718/23, and R6010-23, HKU Seed Fund for Basic Research for New Staff via Project 2201100596.

\bibliography{reference}

\appendix

\subsection{The upper bounds of quantum capacity of \texorpdfstring{$\calO$}{}}
Here we prove the upper bound of quantum capacity for the generalized platypus channel, which is stated in Theorem~\ref{thm1}. Without loss of generality, suppose that $\mu_0 \leq \mu_1 \leq \cdots \leq \mu_{d-1}$. For simply the notation, let $[s] = |s\rangle\langle s|$. Let $|\Phi\rangle = \sum_i |ii\rangle$ is the non-normalized maximum entangled state, the Choi matrix of $\calO$ is 
\begin{equation}
\begin{aligned}
    \mathcal{J}_{\calO} & :=  \mathcal{I}\otimes \calO(|\Phi\rangle\langle\Phi|) \\
    & = \sum_{j=0}^{d-1} \mu_j [0j] + \sum_{i=1}^{d} [i,d] \\
    & \quad + \sum_{i=1}^{d} \sqrt{\mu_{i-1}}\Big(|0,i-1\rangle\langle i,d| + |i,d\rangle\langle 0,i-1|\Big) \,.
\end{aligned}
\end{equation}
We now proof the Theorem~\ref{thm1}.
\begin{proof}
    We use the upper bound given for any quantum channel $\mathcal{B}$~\cite{TranspositionBound}, which is given by
    \begin{equation*}
        Q(\mathcal{B}) \leq \log || \mathcal{T} \circ \mathcal{B}||_{\diamond}
    \end{equation*}
    where $\mathcal{T}$ is the transpose map and $|| \mathcal{T} \circ \mathcal{B}||_{\diamond}$ is the solution of following SDP
    \begin{equation*}
    \begin{aligned}
        || \mathcal{T} \circ \mathcal{B}||_{\diamond} = \min & \ \frac{1}{2}\Big(||Y_{a}||_\infty + || Z_a||_\infty\Big) \\
        & s.t.\, Y_{ab}, Z_{ab} \geq 0 \\
        & \quad \begin{pmatrix}
            Y_{ab} & -\mathcal{T}_b(\mathcal{J}_\N) \\
            -\mathcal{T}_b(\mathcal{J}_\N)^* & Z_{ab}
        \end{pmatrix} \geq 0\,.
    \end{aligned}
    \end{equation*}
    where $\mathcal{T}_b = \mathcal{I}_a\otimes \mathcal{T}$ is partial transpose on the $B$ system and $Y_a, Z_a$ are reduced operators of $Y_{ab}, Z_{ab}$.

    Let $Y_{ab} = Z_{ab}$ are 
    \begin{equation*}
    \begin{aligned}
        Y_{ab}  = \sum_{j=0}^{d-1} \mu_j [0j] + \sum_{i=1}^{d} [i,d] + s [0,d] + |\psi\rangle\langle\psi|
    \end{aligned}
    \end{equation*}
    with $|\psi\rangle = \sum_{i=1}^{d} s_i |i,i-1\rangle$ is non-normalized and all $s, s_j \geq 0$. It's clear that $Y_{ab} \geq 0$, to determine the last matrix in the SDP is positive semi-definite, according to Schur complement, we only need to check
    \begin{equation*}
        Y_{ab} \geq \mathcal{T}_b(\mathcal{J}_\N)\,Y_{ab}^{-1}\, \mathcal{T}_b(\mathcal{J}_\N)\,,\quad (\mathds{1} - Y_{ab}Y_{ab}^{-1}) \mathcal{T}_b(\mathcal{J}_\N) = 0\,,
    \end{equation*}
    where $Y_{ab}^{-1}$ is the Moore-Penrose inverse of $Y_{ab}$, and is given by
    \begin{equation*}
        Y_{ab}^{-1} = \sum_{j=0}^{d-1} \frac{[0j]}{\mu_j}  + \sum_{i=1}^{d} [j,d] + \frac{[0,d]}{s} + \frac{|\psi\rangle\langle\psi|}{(\sum_i s_i^2)^{2}}\,.
    \end{equation*}
    Since 
    \begin{equation*}
        \begin{aligned}
          & \mathcal{T}_b(\mathcal{J}_\N)\,Y_{ab}^{-1}\, \mathcal{T}_b(\mathcal{J}_\N)  \\
          & \quad = \sum_{j=0}^{d-1} \mu_j[0j]  + \sum_{i=1}^{d} [j,d] + \frac{(\sum_i \sqrt{\mu_{i-1}}s_i)^2}{(\sum_i s_i^2)^{2}}[0,d] \\
          & \qquad + \sum_{i,j=1}^{d} \frac{\sqrt{\mu_{i-1}\mu_{j-1}}}{s} |i,i-1\rangle\langle j,j-1|\,, \\
          & (\mathds{1} - Y_{ab}Y_{ab}^{-1}) \mathcal{T}_b(\mathcal{J}_\N)  \\
          & \quad = \sum_{j=1}^{d} \Big(\sqrt{\mu_{j-1}} - \sum_{i=1}^{d} \frac{s_is_j\sqrt{\mu_{i-1}}}{\sum_i s_i^2}\Big) |j,j-1\rangle\langle 0,d|\,.
        \end{aligned}
    \end{equation*}
    we need
    \begin{equation}\label{condition}
    \begin{aligned}
        & \frac{(\sum_i \sqrt{\mu_{i-1}}s_i)^2}{(\sum_i s_i^2)^{2}} \leq s\,. \\
        & \frac{\sqrt{\mu_{i-1}}}{\sqrt{s}} \leq s_i\,, \forall i = 1,\cdots,d\,,\\
        & \sum_{i=1}^{d} \frac{s_is_j\sqrt{\mu_{i-1}}}{\sum_i s_i^2} = \sqrt{\mu_{j-1}}\,, \forall j = 1,\cdots,d\,.
    \end{aligned}
    \end{equation}
    While $Y_a = \tr_b (Y_{ab}) = (1+s) [0] + \sum_{i=1}^{d} (1+s_i^2)[i]$, and $||Y_a||_\infty = \max\{1+s,1+s_1^2,\cdots,1+s_{d}^2\}$. Here we choose 
    \begin{equation*}
        s_i = (\mu_{i-1}^2/\mu_{d-1})^{1/4}\,,
    \end{equation*}
    then \eqref{condition} comes
    \begin{equation*}
        \begin{aligned}
            & \frac{1}{\mu_{d-1}^{1/2}}\bigg/ \frac{1}{\mu_{d-1}} = \mu_{d-1}^{1/2} \leq s\,, \\
            & \mu_{d-1}^{1/2} \leq s\,, \\
            & \frac{\sqrt{\mu_{j-1}}}{\mu_{d-1}^{1/2}}\bigg/ \frac{1}{\mu_{d-1}^{1/2}} = \sqrt{\mu_{j-1}}\,,\, \forall j = 1,\cdots,d\,.
        \end{aligned}
    \end{equation*}
    Thus we choose $s = \mu_{d-1}^{1/2}$, then $||Y_a||_\infty = 1+\sqrt{\mu_{d-1}}$.
\end{proof}

\subsection{Proof of Theorem~\ref{thm2}}

In this section, we firstly use the SDP bound for classical capacity~\cite{SDPBoundC} to evaluate the private and classical capacity of $\calO$.
\begin{proof}
    Due to the inequality $P^{(1)}(\calO) \leq P(\calO) \leq C(\calO)$, we firstly show $P^{(1)}(\calO) \geq 1$.

    Consider the equiprobable ensemble containing two quantum states
    \begin{equation*}
        \rho_0 = [0],\quad \rho_1 = \sum_{i=1}^{d} \mu_{i-1} [i]\,.
    \end{equation*}
    Then the mixture state is $\rho = \frac{1}{2}\Big([0] + \sum_{i=1}^{d} \mu_{i-1} [i]\Big)$, and
    \begin{equation*}
        I_c(\rho,\calO) = \sum_i \frac{\mu_{i-1}}{2}\log \mu_{i-1} + 1\,,
    \end{equation*}
    while
    \begin{equation*}
        I_c(\rho_0,\calO) = 0\,, \qquad I_c(\rho_1,\calO) = \sum_i \mu_{i-1}\log\mu_{i-1}\,.
    \end{equation*}
    Thus $P^{(1)}(\calO) \geq 1\,.$

    We then use the upper bound of classical capacity for any quantum channel $\mathcal{B}$~\cite{SDPBoundC}, which is given by 
    \begin{equation*}
        C(\mathcal{B}) \leq \log \beta(\mathcal{B})\,,
    \end{equation*}
    where $\beta(\mathcal{B})$ is determined by the SDP
    \begin{equation*}
        \begin{aligned}
            \beta(\mathcal{B}) = & \min \tr(S_b) \\
            & \text{s.t.}\, R_{ab}, S_b\, \text{Hermitian} \\
            & \quad - R_{ab} \leq \mathcal{T}_b(\mathcal{J}_\mathcal{B}) \leq R_{ab} \\
            & \quad -\mathds{1}_a\otimes S_b \leq  \mathcal{T}_b(R_{ab}) \leq \mathds{1}_a\otimes S_b\,.
        \end{aligned}
    \end{equation*}
    We consider the $R_{ab}$ and $S_b$ as 
    \begin{equation*}
    \begin{aligned}
         R_{ab} = \sum_{j=0}^{d-1} \mu_j [0,j] + \sum_{i=0}^{d} [i, d] + [\psi]\,,  \quad
         S_b = \sum_{j=0}^{d-1} \mu_j [j] + [d]\,.
    \end{aligned}
    \end{equation*}
    with $|\psi\rangle = \sum_{i=1}^{d} \sqrt{\mu_{i-1}} |i\rangle\otimes|i-1\rangle$. 
    
    It's clear that $-R_{ab} \leq \mathcal{T}_b(\mathcal{J}_{\calO}) \leq R_{ab}$ and $\mathds{1}_a\otimes S_b \pm \mathcal{T}_b(R_{ab}) \geq 0$\,. As $\tr(S_b) = 2$, we have $C(\calO) \leq 1$.
\end{proof}
As for the entanglement-assisted classical capacity of $\calO$, we consider the state $\rho = \frac{1}{2}\Big([0] + \sum_{i=1}^{d} \mu_{i-1} [i]\Big)$ whose purification is $|\psi\rangle = \frac{|00\rangle}{\sqrt{2}} + \frac{1}{\sqrt{2}}(\sum_{i=1}^{d} \mu_{i-1} |ii\rangle)$. Therefore $C_E(\calO) = \max_\rho I(\rho,\calO) \geq  2$. Since the mutual information $I(\rho,\calO)$ is concave on $\rho$, the optimality of $|\psi\rangle$ can be verified by the convex optimization software like CVX and Mosek~\cite{Fawzi_2018}.

\subsection{Super-additivity of quantum capacity of \texorpdfstring{$\calO$ with $\mathcal{E}_{\lambda,d}$}{}}

Now we show the super-additivity of quantum capacity for $\calO$ and the qudit erasure channel $\mathcal{E}_{\lambda,d}$ where $\{\mu_i\}$ to be determined. Here we choose the input state of $\calO\otimes\mathcal{E}_{\lambda,d}$ as
\begin{equation*}
    \rho = \frac{1}{2} [0]\otimes \frac{\mathds{1}_{d}}{d} + \frac{1}{2} |\psi\rangle\langle\psi|
\end{equation*}
with $|\psi\rangle = \sum_{i=1}^{d} \sqrt{\mu_{i-1}} |i\rangle\otimes|i-1\rangle$\,. Thus the output states of $\calO\otimes \mathcal{E}_{\lambda,d}$ and its complementary channel with respect to the input state $\rho$ are
\begin{equation*}
    \begin{aligned}
        & \calO\otimes \mathcal{E}_{\lambda,d}(\rho) \\
        & \, = \frac{1}{2} \sum_{j=0}^{d-1} \mu_{j} [j]\otimes \Big((1-\lambda)\frac{\mathds{1}_{d}}{d}\oplus \lambda [e]\Big) \\
        & \quad + \frac{1}{2}\sum_{i=0}^{d-1}\mu_{i} [d]\otimes \Big((1-\lambda)[i]\oplus \lambda[e]\Big) \,, \\
        & \calO^c\otimes\mathcal{E}_{\lambda,d}^c(\rho) \\
        &\, = \frac{1}{2} \sum_{i=0}^{d-1} \mu_i [i]\otimes\Big(\lambda\frac{\mathds{1}_{d}}{d}\oplus (1-\lambda) [e]\Big) \\
        & \quad + \frac{1}{2} \sum_{i,j=0}^{d-1} \sqrt{\mu_i\mu_j} |i\rangle\langle j|\otimes \Big(\lambda |i\rangle\langle j| \oplus (1-\lambda)\delta_{ij} [e]\Big)
    \end{aligned}    
\end{equation*}
where $\delta_{ij} = 1$ if $i=j$, otherwise $\delta_{ij} = 0$. The eigenvalues of $\calO\otimes \mathcal{E}_{\lambda,d}(\rho)$ are $\frac{\mu_i(1-\lambda)}{2d}$ with multiplicity $d$ and 
$\mu_i\lambda/2\,, \mu_i(1-\lambda)/2\,, \lambda/2$ with multiplicity $1$. The eigenvalues of $\calO^c\otimes\mathcal{E}_{\lambda,d}^c(\rho)$ are more complicate, it's divided into three class
\begin{equation*}
\begin{aligned}
     \calO^c\otimes\mathcal{E}_{\lambda,d}^c(\rho) & = (1-\lambda) \sum_{i=0}^{d-1} \mu_i [i,e] + \frac{\lambda}{2} \sum_{i=0}^{d-1} \mu_i [i]\otimes \frac{\mathds{1}_d-[i]}{d} \\
     & \quad + \frac{\lambda}{2} \Big(\sum_{i=0}^{d-1} \frac{\mu_i}{d} [i,i] + \sum_{i,j=0}^{d-1} \sqrt{\mu_{i}\mu_{j}} |ii\rangle\langle jj| \Big) \\
     & =: (1-\lambda) A_1 + \frac{\lambda}{2} A_2 + \frac{\lambda}{2} A_3\,. 
\end{aligned}
\end{equation*}
These three matrices are orthogonal to each other, therefore the eigenvalues of $\calO^c\otimes\mathcal{E}_{\lambda,d}^c(\rho)$ contain the eigenvalues of first two matrices $(1-\lambda)A_1$ and $\frac{\lambda}{2}A_2$: $\frac{\lambda\mu_i}{2d}$ with multiplicity $d-1$, $(1-
\lambda)\mu_i$ with multiplicity $1$. The eigenvalues of last matrix $\frac{\lambda}{2}A_3$ are 
same as following matrix
\begin{equation}\label{B1}
    B = \frac{\lambda}{2}\begin{pmatrix}
        \frac{(d+1)\mu_0}{d} & \sqrt{\mu_0\mu_1} & \cdots & \sqrt{\mu_0\mu_{d-1}} \\
        \sqrt{\mu_1\mu_0} & \frac{(d+1)\mu_1}{d} & \cdots & \sqrt{\mu_1\mu_{d-1}}  \\
        \vdots & \vdots & \ddots & \vdots \\
        \sqrt{\mu_{d-1}\mu_0} & \sqrt{\mu_{d-1}\mu_1} & \cdots & \frac{(d+1)\mu_{d-1}}{d}
    \end{pmatrix}\,,
\end{equation}
whose eigenvalues are with $d$ non-zero at most, assume as $\frac{\lambda}{2}(\xi_0\,\ldots,\xi_{d-1})$. All these $\xi_i$ satisfy the characteristic equation: 
\begin{equation}\label{CEB}
    \sum_{i=0}^{d-1} \mu_i \Big[\big(x-\frac{\mu_i}{d}-1\big) \prod_{j=0,j\neq i}^{d-1} \big(x-\frac{\mu_j}{d}\big) \Big]= 0
\end{equation}
which can be solved by some numerical methods and software like MATLAB. In particular, for the special probability vector $\vec{\mu}$ such as $\vec{\mu} = (1/d,\cdots,1/d)$, it has analytical expressions. Here we use Weyl inequality to give a bound for the entropy of $B$.

Notice $B$ is the sum of two positive semi-definite matrices $B_1$ and $B_2$ whose eigenvalues are $\lambda(B_1) = \frac{\lambda}{2d}(\mu_0,\ldots,\mu_{d-1})$ and $\lambda(B_2) = (\lambda/2,0,\ldots,0)$, according to the Weyl inequality~\cite{Weylinequality,MatrixAnalysisHorn}, we have 
$$
\frac{\mu_i}{d} \leq \xi_i \leq \frac{\mu_{i+1}}{d}\,,\, \text{for}\\, i = 0,\ldots,d-2\,,\quad \frac{\mu_{d-1}}{d}+1 \leq \xi_{d-1}\,.
$$
Since the entropy is a concave function, using the property of (weak) majorization, we have
\begin{equation*}
    \begin{aligned}
         & I_c\big(\rho, \calO\otimes \mathcal{E}_{\lambda,d}\big) \\
         &  \geq -\sum_{i=0}^{d-1} \frac{\mu_i(1-\lambda)}{2}\log\frac{\mu_i(1-\lambda)}{2d} -\frac{\lambda}{2}\log \frac{\lambda}{2} \\
         &\quad - \sum_{i=0}^{d-1} \Big( \frac{\mu_i\lambda}{2} \log \frac{\mu_i\lambda}{2} + \frac{\mu_i(1-\lambda)}{2}\log \frac{\mu_i(1-\lambda)}{2} \Big)  \\
         &\quad + \sum_{i=0}^{d-1} \Big(\frac{\lambda\mu_i(d-1)}{2d}\log \frac{\lambda\mu_i}{2d} + (1-\lambda)\mu_i \log (1-\lambda)\mu_i \Big) \\
         & \quad + \sum_{i=0}^{d-2} \frac{\lambda\mu_i}{2d}\log \frac{\lambda\mu_i}{2d} + \frac{\lambda}{2}\big(\frac{\mu_{d-1}}{d}+1\big)\log\frac{\lambda}{2}\big(\frac{\mu_{d-1}}{d}+1\big) \\
        & = (1-\lambda) + \frac{d-(2d-\mu_{d-1})\lambda}{2d}\log d\,.
        \end{aligned}
\end{equation*}
Thus
\begin{equation*}
I_c(\rho,\calO\otimes\mathcal{E}_{\lambda,d}) - Q(\mathcal{E}_{\lambda,d}) \geq  1-\lambda + \frac{\mu_{d-1}\lambda-d|1-2\lambda|}{2d}\log d \,.
\end{equation*}
Similar, using the exact expression of coherent information of $\calO$, we can obtain a lower bound for the coherent information $\Q(\calO$. Thus we can obtain the range for $\mu_{d-1}$ and $\lambda$ yielding the super-additivity of quantum capacity and coherent information for the generalized platypus channel $\calO$ and the qudit erasure channel $\mathcal{E}_{\lambda,d}$. 

Moreover, when $\lambda = 1/2$, using our upper bound of $Q(\calO)$, the condition of super-additivity of quantum capacity can be weakened as 
\begin{equation}
    \log(1+\sqrt{\mu_{d-1}}) \leq 1/2 \Longleftrightarrow \mu_{d-1} \leq 3-2\sqrt{2}\,.
\end{equation}
Since $\max_i \mu_i \geq 1/d$, thus $d\geq 3+2\sqrt{2} > 5$. In fact, we can use MATLAB to evaluate the exact eigenvalues of $B$, which gives a better result: for $d>3$, one can find a probability vector such that $Q(\calO\otimes \mathcal{E}_{1/2, d}) > Q(\calO)$, for example, for $d = 4$, $\calO = \M_{5}$ which as shown in~\cite{platypusPRL}.

\subsection{Super-additivity of quantum capacity of \texorpdfstring{$\calO$}{} with multilevel amplitude damping channels}

In this section, we will show the super-additivity of quantum capacity for $\calO$ associated with the multilevel amplitude damping channel. We consider a special amplitude damping channel $\mathcal{A}_\gamma$, whose Kraus operators are:
\begin{equation}\label{KrausAD}
    \begin{aligned}
    K_{0} & = |0\rangle\langle 0| + \sum_{j=1}^{d-2}\sqrt{1-\gamma} |j\rangle\langle j| \,, \\
    K_{j} & = \sqrt{\gamma} |0\rangle \langle j| \,,\, \text{for}\, j = 1,\cdots,d-1\,. 
    \end{aligned}
\end{equation}
the output states of the complementary channel $\mathcal{A}_{1-\gamma}$ with the input state $\sigma = \sum_{ij} \sigma_{ij} |i\rangle\langle j|$ are 
\begin{equation*}
    \mathcal{A}_{1-\gamma}(\sigma) = \begin{pmatrix}
        \gamma \sigma_{00} + 1-\gamma & \sqrt{\gamma} \sigma_{01} & \cdots & \sqrt{\gamma}\sigma_{0,d-1} \\
        \sqrt{\gamma}\sigma_{10} & \gamma \sigma_{11} & \cdots & \gamma \sigma_{1,d-1} \\
        \vdots &  \vdots & \ddots & \vdots \\
        \sqrt{\gamma}\sigma_{d-1,0} & \gamma\sigma_{d-1,1} & \cdots & \gamma \sigma_{d-1,d-1}
    \end{pmatrix}.
\end{equation*}
The complementary channel $\mathcal{A}^c$ is also a special amplitude damping channel with error probability $1-\gamma$, thus making $\mathcal{A}_{\gamma}$ is degradable for $\gamma\in [0,1/2]$ and anti-degradable for $\gamma\in [1/2,1]$. Using the covariant property of multi-level amplitude damping, the quantum and private capacity is attained on the diagonal state~\cite{GADC1,GADC2}:
\begin{equation*}
     Q(\mathcal{A}_\gamma) = \Q(\mathcal{A}_\gamma) = \max_{\sigma_{\text{diag}}} I_c(\sigma_{\text{diag}},\mathcal{A}_\gamma) = \max_{\sigma(u)}I_c(\sigma(u),\mathcal{A}_\gamma)
\end{equation*}
where $\sigma(u) = \text{diag}(u,\frac{1-u}{d-1},\cdots,\frac{1-u}{d-1})$.

Inputting the state $\rho$ in Eq.~\eqref{state} into $\calO\otimes\mathcal{A}$ and its complement, let $[j]:= |j\rangle\langle j|$ and $\text{diag}(\vec{a})$ is a diagonal matrix with the entry of $\vec{a}$ in its diagonal part for simple, the output states are 
\begin{equation*}
    \begin{aligned}
        & \calO\otimes \mathcal{A}_\gamma(\rho) \\
        & = \frac{1}{2}\sum_{j=0}^{d-1} \mu_j [j]\otimes \frac{1}{d} \Big(\big(1+(d-1)\gamma\big) [0] + (1-\gamma) \sum_{i=1}^{d-1} [i]\Big) \\
        & \quad + \frac{1}{2} [d]\otimes \Big(\big(\mu_0 + \gamma(1-\mu_0)\big)[0] + (1-\gamma) \sum_{i=1}^{d-1} \mu_i [i]\Big)\,, \\
        & \calO^c\otimes\mathcal{A}^c_\gamma(\rho) \\
         & = \frac{1}{2}\sum_{j=0}^{d-1} \mu_j [j]\otimes \frac{1}{d} \Big(\big(1+(d-1)(1-\gamma)\big) [0] + \gamma \sum_{i=1}^{d-1} [i]\Big) \\
        & \quad + \frac{1}{2}\bigg[(1-\gamma)\sum_{i=1}^{d-1} \mu_i [i,0] + [\psi]\bigg]\,,
    \end{aligned}
\end{equation*}
where $\psi = \sqrt{\mu_0} |00\rangle + \sum_{i=1}^{d-1} \sqrt{\mu_i \gamma}|ii\rangle $ is a non-normalized quantum state. Since the eigenvalues of $ \calO\otimes \mathcal{A}_\gamma(\rho)$ is simple, its entropy can be evaluated directly.

While the eigenvalues of $\calO^c\otimes \mathcal{A}^c_\gamma(\rho)$ are $\frac{\mu_0\gamma}{2d}$ with multiplicity $d-1$, $\frac{\mu_j(1+(2d-1)(1-\gamma)}{2d}$ with multiplicity $1$, $\frac{\gamma\mu_{j}}{2d}$ with multiplicity $d-2$, for $j = 1,\cdots,d-1$, as well as the eigenvalues of following matrix
\begin{equation*}
    A = \begin{pmatrix}
        \frac{((d+1)+(d-1)(1-\gamma))\mu_0}{2d} & \frac{\sqrt{\mu_0\mu_1\gamma}}{2} & \cdots & \frac{\sqrt{\mu_0\mu_{d-1}\gamma}}{2}\\
        \frac{\sqrt{\mu_0\mu_1\gamma}}{2} & \frac{(d+1)\gamma\mu_1}{2d} & \cdots  & \frac{\gamma\sqrt{\mu_1\mu_{d-1}}}{2} \\
        \vdots &  \vdots & \ddots & \vdots \\
        \frac{\sqrt{\mu_0\mu_{d-1}\gamma}}{2} & \frac{\gamma\sqrt{\mu_{d-1}\mu_1}}{2} & \cdots & \frac{(d+1)\gamma\mu_{d-1}}{2d}
    \end{pmatrix}\,.
\end{equation*}
the eigenvalues of $A$ satisfy the similar but more complicate form of the matrix in~\eqref{B1} $B$'s characteristic equation as Eq.~\eqref{CEB}, which can also be solved by MATLAB. Here we also give a bound for the entropy of $A$.
% $$
% \begin{aligned}
%   \prod_{i=1}^{d-1}\Big(x-\frac{\gamma\mu_i}{2d}\Big)\Big(x-\frac{2d-(d-1)\gamma}{2d}\mu_0\Big) - \Big(x-\frac{d-(d-1)\gamma}{2d}\mu_0\Big)\sum_{i=1}^{d-1} \frac{\mu_i \gamma}{2} \prod_{j=1,j\neq i}^{d-1} \big(x - \frac{\mu_j\gamma}{2d}\big)
% \end{aligned}
% $$
The Weyl inequality also suits this matrix and can give an upper bound for its entropy. Here we use another majorization inequality~\cite{Rotfel’d1969,Thompson1977,Marshall1980}: $\big(\lambda(B_2+B_1),0\big) \succ \big(\lambda(B_2), \lambda(B_1)\big)$, and notice the matrix $A$ is the sum of two positive semi-definite matrices, whose eigenvalues are $\lambda(B_2) = (\frac{(1+(d-1)\gamma)\mu_0}{2d}, \frac{\gamma\mu_{1}}{2d},\ldots,\frac{\gamma\mu_{d-1}}{2d})$ and $ \lambda(B_1) = (\frac{\gamma(1-\mu_0) + \mu_0}{2},0,\cdots,0)$, we have 
\begin{equation*}
    \begin{aligned}
        & I_c(\rho,\calO\otimes \mathcal{A}_\gamma) \\
        & \geq S\Big(\calO\otimes \mathcal{A}_\gamma(\rho)\Big) + \frac{(d-1)\mu_0\gamma}{2d} \log \frac{\mu_0\gamma}{2d} \\
        & + \sum_{j=1}^{d-1} \frac{\mu_j(1+(2d-1)(1-\gamma))}{2d} \log \frac{\mu_j(1+(2d-1)(1-\gamma))}{2d} \\
        & + \sum_{j=1}^{d-1} \frac{(d-2)\mu_j\gamma}{2d}\log \frac{\mu_j\gamma}{2d} - S(B_1) - S(B_2)
    \end{aligned}
\end{equation*}
It's hard to give an analytical result about the exact value of above expression, but after the detailed computation  and using the simple inequality $\mu_{d-1} \leq 1-(d-1)\mu_0$, such lower bound can be shown only relying on $\mu_{d-1}$ and $\gamma$ for fixed $d$, we display the range of $\mu_{d-1}$ and $\gamma$ for different $d$ such that $\calO$ has super-additive quantum capacity in Fig.~\ref{fig3}(a),(b). Following, we deduce analytical results for some special situations.

The first case: assuming $\gamma = 1/2$, hence we have
\begin{equation}
\begin{aligned}
    & I_c(\rho,\calO\otimes\mathcal{A}_{1/2}) \\
    & \geq (1-\mu_0)\Big(\frac{1}{4} \log \frac{2d+1}{d} + \frac{d+1}{4d} \log \frac{2d+1}{d+1}\Big) \\
    & \geq \frac{1-\mu_0}{2}\,.
\end{aligned}
\end{equation}
Since the quantum capacity of $\mathcal{A}_{1/2}$ is zero, and the upper bound of quantum capacity of $\calO$ is 
$$\log (1+\sqrt{\mu_{d-1})} \leq \sqrt{\mu_{d-1}} \leq \sqrt{1-(d-1)\mu_0}\,,$$ 
together with $\mu_0 \leq 1/d$, one can find when $d\geq 6$, there exists a probability vector $\vec{\mu}$ so that $$Q(\calO\otimes \mathcal{A}_{1/2}) \geq Q(\calO).$$

The second case: suppose that $\mu_i = 1/d$ for all $i$, that is $\calO = \M_{d+1}$, the eigenvalues of the matrix $A$ can be computed in detail, which are $\frac{\gamma}{2d^2}$ with multiplicity $d-2$, and $x_\pm = \frac{2d + (d^2-2d+2)\gamma \pm d\sqrt{d^2\gamma^2 - 4\gamma +4}}{4d^2}$. Therefore, the term $I_c(\rho,\M_{d+1}\otimes\mathcal{A}_{1/2})$ only depends on $\gamma$, and we show the range of $\gamma$ such that $\M_{d+1}$ has super-additive quantum capacity in Fig.~\ref{fig3}(c).
% \begin{equation}
% \begin{aligned}
%     I_c(\rho,\M_{d+1}\otimes\mathcal{A}_\gamma) & = \frac{1}{2} \bigg[\log d - \frac{1+(d-1)\gamma}{d}\log\frac{1+(d-1)\gamma}{d} - (d-1)\frac{1-\gamma}{d}\log \frac{1-\gamma}{d}\bigg] \\
%     & \quad - \frac{1}{2} \bigg[\frac{1+(d-1)\gamma}{d}\log \frac{1+(d-1)\gamma}{d} +  \frac{(d-1)(1-\gamma}{d}\log \frac{1-\gamma}{d} \bigg] + 1 \\
%     & \quad + \frac{(d-1)\gamma}{2d^2} \log \frac{\gamma}{2d^2} + \frac{(d-1)(1+(2d-1)(1-\gamma))}{2d^2} \log \frac{(1+(2d-1)(1-\gamma))}{2d^2} + \frac{(d-1)(d-2)\gamma}{2d^2}\log \frac{\gamma}{2d^2} \\
%     & \quad + \frac{(d-2)\gamma}{2d^2}\log \frac{\gamma}{2d^2} + x_+\log x_+ + x_-\log x_- \\
% \end{aligned}
% \end{equation}

\end{document}